\documentclass[twocolumn,pra,amsmath,amsfonts,amssymb,superscriptaddress]{revtex4-1}

\pdfoutput=1

\usepackage{bm,color}
\usepackage{graphicx}

\newcommand{\su}[1]{\mathrm{SU}(#1)}
\newcommand{\ket}[1]{\left\vert#1\right\rangle}
\newcommand{\bra}[1]{\left\langle#1\right\vert}
\newcommand{\tb}{\bar{\bm{3}}}

\begin{document}

\title{A gapped $\mathrm{SU}(3)$ spin liquid with $\mathbb Z_3$ topological order}
\author{Ivana Kure\v{c}i\'c}
\affiliation{Max-Planck-Institute of Quantum Optics, Hans-Kopfermann-Str.\
1, 85748 Garching, Germany}
\author{Laurens Vanderstraeten}
\affiliation{Department of Physics and Astronomy, Ghent University,
Krijgslaan 281, S9, 9000 Gent, Belgium}
\author{Norbert Schuch}
\affiliation{Max-Planck-Institute of Quantum Optics, Hans-Kopfermann-Str.\
1, 85748 Garching, Germany}

\begin{abstract}
We construct a topological spin liquid (TSL) model on the kagome lattice, with
$\mathrm{SU}(3)$ symmetry with the fundamental representation at each
lattice site, based on Projected Entangled Pair States (PEPS).  Using the
PEPS framework, we can adiabatically connect the model to a fixed point model
(analogous to the dimer model for Resonating Valence Bond states) which we
prove to be locally equivalent to a $\mathbb Z_3$ quantum double model.
Numerical study of the interpolation reveals no sign of a phase transition
or long-range order, characterizing the model conclusively as a gapped TSL. We
further study the entanglement spectrum of the model and find that while
it is gapped, it exhibits branches with vastly different velocities, with
the slow branch matching the counting of a chiral $\mathrm{SU}(3)_1$ CFT,
suggesting that it can be deformed to a model with chiral
$\mathrm{SU}(3)_1$ entanglement spectrum.
\end{abstract}
 
\maketitle

\section{Introduction}

Topological spin liquids (TSL) are exotic phases of matter where a
two-dimensional 
quantum magnet orders topologically rather than magnetically. Identifying
TSLs is subtle, as it requires not only to characterize the topological
order but also to certify the absence of any kind of conventional
long-range order. An important tool in the theoretical study of TSLs is
the use of variational wavefunctions, which allow for a more direct
encoding of desired properties and can be more succinctly analyzed. The
best known candidate for a TSL wavefunction is the $\mathrm{SU}(2)$
Resonating Valence Bond (RVB) state~\cite{anderson:rvb-highTC} on the
kagome or other frustrated lattices.  Its understanding has been
significantly advanced by considering quantum dimer models, where singlets
are replaced by
orthogonal dimers~\cite{moessner:dimer-triangular,misguich:dimer-kagome}, but their precise relation to the $\mathrm{SU}(2)$-invariant RVB state
has long been open.  In recent years, Projected Entangled Pair States
(PEPS)~\cite{verstraete:mbc-peps} have been established as a
tool to study spin liquid wavefunctions. In particular, they could be
used to provide a unified description of the RVB and the quantum dimer
wavefunction, thereby allowing to interpolate between them and to
unambiguously identify the topological phase; PEPS-specific transfer
operator techniques additionally allowed to certify the absence of
any kind of conventional order in the spin degrees of
freedom~\cite{schuch:rvb-kagome}. 

A key insight that is provided by these PEPS constructions concerns the
interplay between the physical symmetry ($\mathrm{SU}(2)$ for the RVB
state) and the nature of the topological phase (here, a $\mathbb Z_2$
Toric Code). Indeed, the fact that PEPS encode properties of a
wavefunction locally makes them natural tools to
study this question: For the RVB, the local encoding of the physical spin
$\tfrac12$ symmetry gives rise to a $\mathbb Z_2$ symmetry in the
entanglement degrees of freedom~\cite{hackenbroich:chiral-su2}, which is
known to underlie topological order and allows for the direct study of
its ground space and topological excitations~\cite{schuch:peps-sym}.
Noteworthily, the $\mathbb Z_2$ order must be of the Toric Code type:
attempts to construct RVB wavefunctions~\cite{iqbal:semionic-rvb} (and
dimer models~\cite{qi:semionic-dimer-models,iqbal:semionic-rvb,buerschaper:semionic-dimermodel})
with Double Semion order (a \emph{twisted} $\mathbb Z_2$ model) break the
lattice symmetry (arising from a symmetry breaking in the mapping from the
dimer model to the topological loop gas), and a no-go theorem proving an
obstruction has been found~\cite{zaletel:su2-semion}, relating to the
half-integer spin per unit cell and the resulting odd $\mathbb Z_2$ parity
in the entanglement. 
In light of this connection between physical symmetry and topological
order it is natural to apply the PEPS framework to TSLs with other
physical symmetry groups, and investigate whether these systems display a
similar interplay between local symmetries and topological order. The most
promising direction is towards spin systems with higher $\mathrm{SU}(N)$
symmetries, especially since these systems have been realized in recent
experiments with ultracold atoms in optical
lattices~\cite{gorshkov:suN-optical-lattice,scazza:suN-optical-lattice,zhang:suN-experiment}
and theoretical and
numerical studies have shown that they potentially host spin-liquid
phases~\cite{hermele:suN-liquids,nataf:suN-chiral-SL}.

In this paper,
we construct a $\mathrm{SU}(3)$ spin liquid wavefunction with $\mathbb
Z_3$ topological order on the kagome lattice.  Specifically, we construct
a PEPS wavefunction with the following properties: (\emph{i}) It has
$\mathrm{SU}(3)$ symmetry with the fundamental representation on each
site, it is invariant under translation and lattice rotations, and
transforms as $\ket\psi\to\ket{\bar\psi}$ under reflection.  (\emph{ii})
The model is a spin liquid, i.e., it exhibits no conventional long-range
order of any kind. (\emph{iii}) The model has topological order and
anyonic excitations corresponding to the $\mathbb Z_3$ quantum double
model, arising from the conservation of $\su3$ ``color'' charge.
(\emph{iv}) The wavefunction is the ground state of a local parent
Hamiltonian and both the wavefunction and the Hamiltonian can be smoothly
connected to a fixed point model with $\mathbb Z_3$ topological order, in
close analogy with the RVB--dimer connection.  (\emph{v}) The
model has trivial charge per unit cell (corresponding to an unbiased
mapping to the topological model), allowing for the possibility to
construct a ``twisted'' version of it.

A natural question given the ``chiral'' transformation property
$\ket\psi\to\ket{\bar\psi}$ under reflection is whether the entanglement
spectrum (ES) exhibits chiral features. We find that the 
ES indeed exhibits left- and right-propagating modes 
with very different velocities, and the slow mode in the trivial
sector displays a level counting clearly matching that of a chiral
$\mathrm{SU}(3)_1$ CFT. Yet, we find clear evidence that the modes couple and
the ES is gapped. However, under a specific deformation
the chiral features become more pronounced, and it is well conceivable
that the ES becomes chiral for instance as the
deformation drives the system through a phase transition.

\section{The $\mathrm{SU}(3)$ model}

\begin{figure}
\includegraphics[width=0.95\columnwidth]{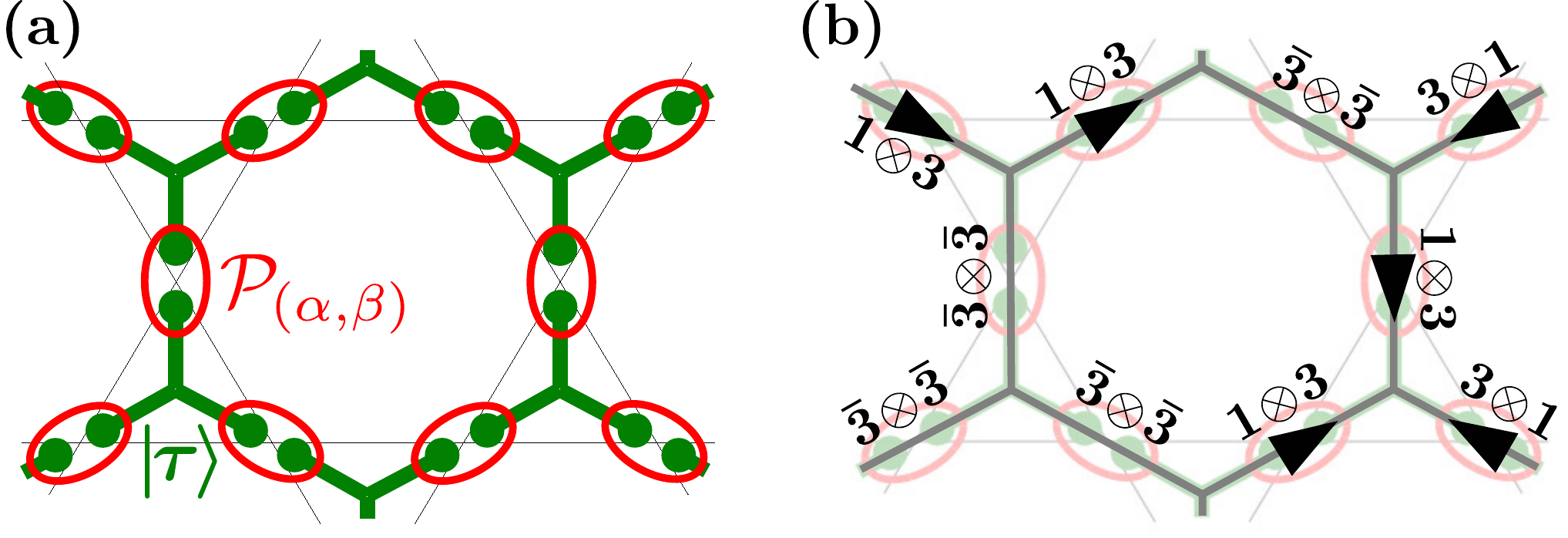}
\caption{\textbf{(a)} The model is constructed from trimers $\ket\tau$
which are in a singlet state with representation
$\mathcal H_v\equiv \bm1\oplus\bm3\oplus\tb$ at each site (green dots), 
to which a map $\mathcal P_\bullet$ is applied which selects the physical
degrees of freedom from
$\mathcal H_v\otimes\mathcal H_v$. \textbf{(b)} Mapping to a $\mathbb Z_3$
topological model: Each site holds a $\mathbb Z_3$ degree of freedom: one
of two arrows or no arrow.  The arrows are pointing towards the $\bm 3$
representation and satisfy a Gauss law across each vertex due to the
fusion rules of the $\mathrm{SU}(3)$ irreps.
}
\label{fig:construction}
\end{figure}

We start by providing the construction of the model, illustrated in
Fig.~\ref{fig:construction}a. 
We start from trimers $\ket\tau$ built of three ``virtual'' particles,
$\ket\tau\in\mathcal H_v^{\otimes 3}$, where each of the virtual particles
lives in $\mathcal H_v=\bm 1 \oplus\bm3\oplus\bar{\bm
3}$.  Here, the boldface numbers denote representations of $\su3$, this
is, $\mathcal H_v$ decays into a direct sum $\mathbb C^1\oplus \mathbb
C^3\oplus\mathbb C^3$, with $u\in\su3$ acting with the trivial action
($\bm 1$), the fundamental action $u$ ($\bm 3$) and the anti-fundamental
action $\bar u$ ($\tb$), respectively. We will choose $\ket\tau$
to be an $\mathrm{SU}(3)$ singlet.  $\mathcal H_v^{\otimes 3}$ supports a
total of $9$ singlets, namely one in each of the spaces
$\bm1\otimes\bm1\otimes\bm1$, 
$\bm 3\otimes \bm3\otimes \bm3$,
and $\tb\otimes\tb\otimes\tb$,
and the 6 permutations of $\bm1\otimes\bm3\otimes\tb$.
We
choose $\ket\tau$ to be an equal weight superposition of all singlets,
with the following convention: The $6$ states 
$\bm1\otimes\bm3\otimes\tb$ together with the $\bm1\otimes\bm1\otimes\bm1$
singlet are combined with amplitudes $\pm1$ to form a fully symmetric
state $\ket S$, the remaining states 
$\bm3\otimes\bm3\otimes \bm3$ and $\tb\otimes\tb\otimes\tb$ are combined with
amplitudes $+1$ to form a fully antisymmetric state $\ket A$, and
$\ket\tau =\ket S + i\ket A$.  The state $\ket\tau$ thus has a
chiral symmetry: It transforms trivially under translation and rotation and as
$\ket\tau\mapsto\ket{\bar\tau}$ under reflection.

We now arrange the trimers $\ket\tau$ as shown in
Fig.~\ref{fig:construction}a and apply maps $\mathcal P_{(\alpha,\beta)}$ to
pairs of adjacent virtual sites, where the parameters
$\alpha,\beta\in[0,1]$
will allow us to interpolate between the fixed point model and the
$\mathrm{SU}(3)$ spin liquid.  We first define the map $\mathcal P_{\perp}\equiv
\mathcal P_{(1,1)}$ which projects the two adjacent sites $\mathcal H_v^{\otimes
2}=(\bm1\oplus\bm3\oplus\tb)^{\otimes 2}$ onto the union of the three
components $\mathcal H_\omega=\bm1\otimes\bm3$, $\mathcal
H_{\bar\omega}=\bm3\otimes\bm1$, and
$\mathcal H_1=\tb\otimes\tb$. We will show in a moment that the resulting
wavefunction is a fixed point wavefunction with $\mathbb Z_3$ topological order.

The interpolation in $\alpha$ is now obtained by adiabatically removing
the $\bar{\bm 6}$ component in $\mathcal H_1=\tb\otimes\tb=\bm
3\oplus\bar{\bm 6}$, that is,
\begin{equation}
    \label{eq:first-P-interpolation}
\mathcal P_{(\alpha,1)}= \Big[\openone_{\mathcal H_\omega}
    \oplus \openone_{\mathcal H_{\bar\omega}}
    \oplus \big(\alpha \openone_{\mathcal H_1} + (1-\alpha) \Pi_{\mathcal
H_1^{\bm 3}}\big)\Big]\, P_\perp\ ,
\end{equation}
where we have decomposed $\mathcal H_1=\bm 3\oplus\bar{\bm 6}\equiv
\mathcal H_1^{\bm 3}\oplus \mathcal H_1^{\bar{\bm{6}}}$, and $\Pi_\mathcal
H$ denotes the orthogonal projector onto $\mathcal H$.

At $\alpha=0$, we are left with $\mathcal P_{\bm{333}}=\mathcal P_{(0,1)}$
which maps into $\mathcal H_{\bm{333}}=\mathcal H_\omega \oplus \mathcal
H_{\bar\omega}\oplus \mathcal H_{1}^{\bm 3}\cong \bm 3 \otimes \mathbb
C^3$, where the first  tensor component transforms 
as $\bm 3$, while the second component labels \emph{which} 
representation we consider, and thus transforms trivially under
$\mathrm{SU}(3)$. We can now remove the $\mathbb C^3$
adiabatically,
\begin{equation}
    \label{eq:second-P-interpolation}
\mathcal P_{(0,\beta)} = \big[\openone_{\bm 3}\otimes 
(\beta\,\openone_{\mathbb C^3}+ (1-\beta)\ket{e}\bra{e}_{\mathbb C^3} 
)\big]\,\mathcal P_{(0,1)}\ ,
\end{equation}
by projecting the label onto the equal weight superposition $\ket{e}$ 
of the three components (with the phases of
$\mathcal H_\omega$, $\mathcal H_{\bar\omega}$ chosen opposite, 
leaving $\mathcal P_{(0,\beta)}$ antisymmetric).
For $(\alpha,\beta)=(0,0)$, we can factor out the $\ket{e}$ and are thus
left with an $\mathrm{SU}(3)$-invariant wavefunction (as the building
blocks are
$\mathrm{SU}(3)$-invariant) with the fundamental representation at each
site.  Clearly, the two interpolations can be combined into a two-parameter
family, though in the following we will only consider the presented
sequence of interpolations $(1,1)\rightarrow(0,1)\rightarrow(0,0)$.

Let us now show how to map the model with $\mathcal P_{(1,1)}=\mathcal
P_\perp:\mathcal H_v^{\otimes 2}\rightarrow \mathcal H_\omega \oplus
\mathcal H_{\bar\omega}\oplus\mathcal H_1$ to a topological
$\mathbb Z_3$ fixed point model by local unitaries.  To this end, let us
first add an extra qutrit (``indicator'')
$\{\ket{-}_t,\ket{\rightarrow}_t,\ket{\leftarrow}_t\}$ at each vertex, onto
which we copy the information whether the system at that vertex is in the
space $\mathcal H_1=\tb\otimes\tb$, $\mathcal H_\omega=\bm1\otimes\bm3$,
or $\mathcal H_{\bar\omega}=\bm3\otimes\bm1$, as shown in
Fig.~\ref{fig:construction}b (the arrow always points towards the $\bm 3$
irrep).  Now consider for a  moment the scenario where we project all
indicator qutrits onto this basis (``classical configurations'').  Given
any such classical configuration, the states of the virtual system
factorize into singlet states of the corresponding irreps on the
individual triangles, and can thus be brought into a fiducial state
by local unitaries controlled by the state of the indicator qutrits, and
thus effectively removed.  

This operation can be done coherently, leaving us with the indicator
qutrits in a superposition of all allowed configurations.
The construction of $\ket\tau$ ensures that around each vertex of the dual
honeycomb lattice, the number of $\bm 3$ (i.e., ingoing arrows) minus the
number of $\bm 1$ (i.e., outgoing arrows) is $0\, \mathrm{mod}\,3$. 
Associating the indicator qutrits with $\mathbb Z_3$
variables (with arrows pointing from the A to the B sublattice
corresponding to $\omega=e^{2\pi i/3}$, the other arrows to $\bar\omega$,
and ``no arrow'' to $1$), it follows that the indicator qutrits which live
on the edges of the honeycomb lattice satisfy a $\mathbb Z_3$ Gauss' law.
In addition, all allowed configurations appear with equal weight.
Thus, the wavefunction given by the indicator qutrits is
the wavefunction of a quantum double model $D(\mathbb Z_3)$, i.e., a fixed
point wavefunction with $\mathbb Z_3$ topological order~\footnote{Observe that
our model is not a ``resonating trimer state'' as e.g.\ in
Ref.~\cite{lee:resonating-trimer-state}, since projecting
$\tb\otimes\tb$ creates large entangled clusters (e.g.\ for the
vaccuum of the $\mathbb Z_3$ model). Placing the model of
Ref.~\cite{lee:resonating-trimer-state} on the kagome lattice in fact
yields a variant of our model with a modified \text{$\vert\tau\rangle$} which we
found to be in a trivial phase.}.

Having established that the model with $\mathcal P_\perp$ is a $\mathbb
Z_3$ topological fixed point model without any long-range order in the
spin degree of freedom (which we could factor out), we can now use the
interpolation $\mathcal P_{(\alpha,\beta)}$ to study whether the topological order and
the spin liquid behavior remain present all the way down to the $\su3$
point. An important point is that along this interpolation, there exists a
corresponding path of parent Hamiltonians.  Continuity of the 
Hamiltonian, as well as the $9$-fold degenerate ground space in the finite
volume (where the $9$ linearly independent ground states are obtained by
placing strings of symmetry actions when closing the boundary),
can be established by showing that the PEPS has a property known as
$G$-injectivity~\cite{schuch:peps-sym} with $G\equiv\mathbb Z_3$,
which states that (\emph{i}) the tensors have a virtual
$\mathbb Z_3$ symmetry -- in
our case, this follows directly from the fact that the number of $\bm
1$ minus the number of $\bm 3$ of both $\ket\tau$ and $\mathcal P_\bullet$
are $0\,\mathrm{mod}\,3$ -- and (\emph{ii}) there exists a blocking where the
mapping $M$ from the virtual degrees of freedom at the boundary to the
degrees of freedom in the bulk given by a patch of the PEPS is injective
on the $\mathbb Z_3$-invariant subspace. We have verified that
(\emph{ii}) holds on a full star using full diagonalization of
$M^\dagger M$~\footnote{Using $\mathrm{U}(1)$ conservation and rewriting
$M^\dagger M$ as a product of $6$ matrices each
depending on the virtual ket+bra indices on one tip of the star; note that
it is sufficient to check this for $\mathcal P_{(0,1)}$, since for
$\beta>0$, $\mathcal P_{(\alpha,\beta)}$ can be mapped to $\mathcal P_{(0,1)}$ by
acting on the physical system.} for all $(\alpha,\beta)$
except for $(0,0)$~\footnote{The latter point might still be $\mathbb
Z_3$-injective, but this would require checking larger patches which seems
numerically prohibitive.}.
This implies the existence of a parent Hamiltonian acting on two
overlapping stars, 
which changes continuously along the interpolation and has the correct
$9$-fold degenerate ground space
structure~\cite[App.~D]{schuch:rvb-kagome}.
For the missing point
$(\alpha,\beta)=(0,0)$, a Hamiltonian can be constructed by taking the limit
of the Hamiltonians for $\alpha=0$, $\beta\rightarrow0$, which is
continuous by construction~\footnote{The existence of the limit
is guaranteed by the fact that the parent Hamiltonian is constructed as
the projector onto the orthocomplement of the image of the PEPS on two
overlapping stars, which is a polynomial (and thus analytic) map in
$\beta$, together with the fact that eigenspace projectors of analytic hermitian
matrices are analytic themselves
(\cite[Thm.~6.1]{kato:pert-of-linear-operators}).}.

\begin{figure}[t]
\includegraphics[width=\columnwidth]{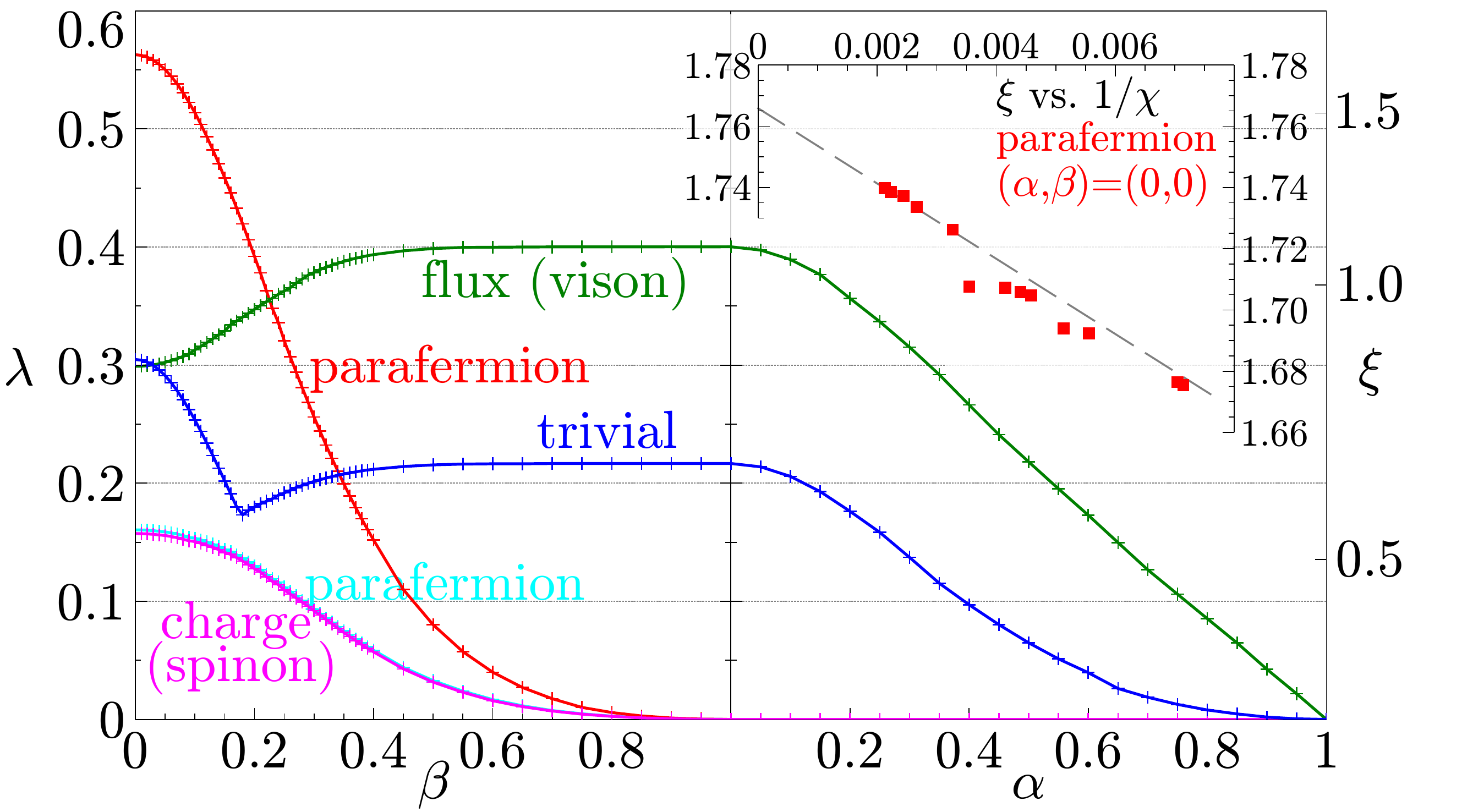}
\caption{
Anyon-anyon and trivial correlations along the interpolation
$(0,0)\leftarrow(0,1)\leftarrow(1,1)$, cf.~text, with $\xi$ the
correlation length and $\lambda=\exp(-1/\xi)$. Data has been obtained
with bond dimensions up to $\chi\sim 500$ and truncating Schmidt values down
to $10^{-16}$.  The data clearly shows that the $\mathrm{SU}(3)$ model is
in the $D(\mathbb Z_3)$ double phase. The inset shows the convergence
of $\xi$ for the parafermion (red) at $(\alpha,\beta)=(0,0)$ as a function
of the inverse bond dimension $1/\chi$; for large $\chi$, $\xi$ exhibits the
linear scaling commonly found for PEPS in this extrapolation, yielding an
approximate value of $\xi\approx1.77$.}
\label{fig:corrlength}
\end{figure}

\section{Numerical study}

Let us now study the behavior along the interpolation.
To this end, we use infinite MPS (iMPS) with tunable bond dimension $\chi$
to approximate the fixed point of the PEPS transfer
operator~\cite{haegeman:medley}.  From the iMPS, we can obtain correlation
lengths $\xi_{g,\eta}$ for general anyon-anyon correlations through the
subleading eigenvalues $\lambda_{g,\eta}=\exp[-1/\xi_{g,\eta}]$ of the
(dressed) iMPS transfer operator~\cite{iqbal:z4-phasetrans}. Here, group
elements $g$ label anyon fluxes, and irreps $\eta$ label anyon charges.
These describe different types of excitations: Fluxes or visons correspond
to an incorrect relative weight of different singlet configurations,
and charges or spinons to a spinful excitation (i.e., a breaking of
singlets); they are modeled by strings of symmetry actions and irrep
actions on the virtual level, respectively. 
Composite
particles of visons and spinons possess a parafermionic $e^{\pm2\pi i/3}$
statistics.  The numerical results for these quantities along the
interpolation are shown in Fig.~\ref{fig:corrlength}.  We find that for
the first part of the interpolation, $(1,1)\to(0,1)$, only vison
excitations acquire a finite correlation length. This can be understood
from the fact that projecting out the $\bar{\bm 6}$ component in
$\tb\otimes\tb=\bm 3\oplus\bar{\bm 6}$ decreases the effective amplitude of
singlets with $\tb$ representations.  In fact, these correlations can be mostly
supressed by considering a modified model where $\mathcal P_{(\alpha,1)}$,
Eq.~(\ref{eq:first-P-interpolation}), carries an additional 
factor in front of the $(1-\alpha)$ term to keep the total weight of the
$\tb\otimes \tb$ subspace fixed;
we term this the $\tb$-enhanced model 
and discuss it in more detail in the Appendix.  
Along the second
interpolation $(0,1)\to(0,0)$, we observe that the vison length does no
longer grow (and ultimately even decreases), while
the length scale towards the $\mathrm{SU}(3)$ point is dominated by one of
the parafermions. At the $\mathrm{SU}(3)$
point, we find $\xi\approx 1.77$ by extrapolating in $1/\chi$~\footnote{The
longest-ranging two-point correlations (corresponding to the trivial
sector) with $\xi\approx0.86$ are obtained for spin-spin correlations.}.
Most importantly, the extrapolation clearly shows that the correlation
length for $\chi\to\infty$ is \emph{finite}, even if one might prefer to 
take the precise extrapolated value with a bit of care.  
The fact that there is no divergence in any correlation thus proves that
the system is gapped, and in the same phase as the $\mathbb Z_3$ quantum
double model, with no anyons
condensed~\cite{duivenvoorden:anyon-condensation,iqbal:z4-phasetrans}.
This is
confirmed by the convergence analysis shown in the inset, and is
consistent with a number of other checks~\footnote{Exact
diagonalization of the transfer operator on small cylinders, the
convergence of the iMPS method, and the smooth change of the
leading eigenvalue and eigenvector of the transfer operator along the
interpolation.}.  For the $\tb$-enhanced model (Fig.~\ref{fig:W3b}a), we
find a strongly decreased correlation length along
the $\alpha$-interpolation, as expected.  Surprisingly, this effect
is reversed along the $\beta$-interpolation, and at the $\su3$
point, $\xi\approx2.64$.  Nevertheless, the results conclusively show that 
this model is in the $D(\mathbb Z_3)$ phase as well.

\begin{figure}[t]
\includegraphics[width=.95\columnwidth]{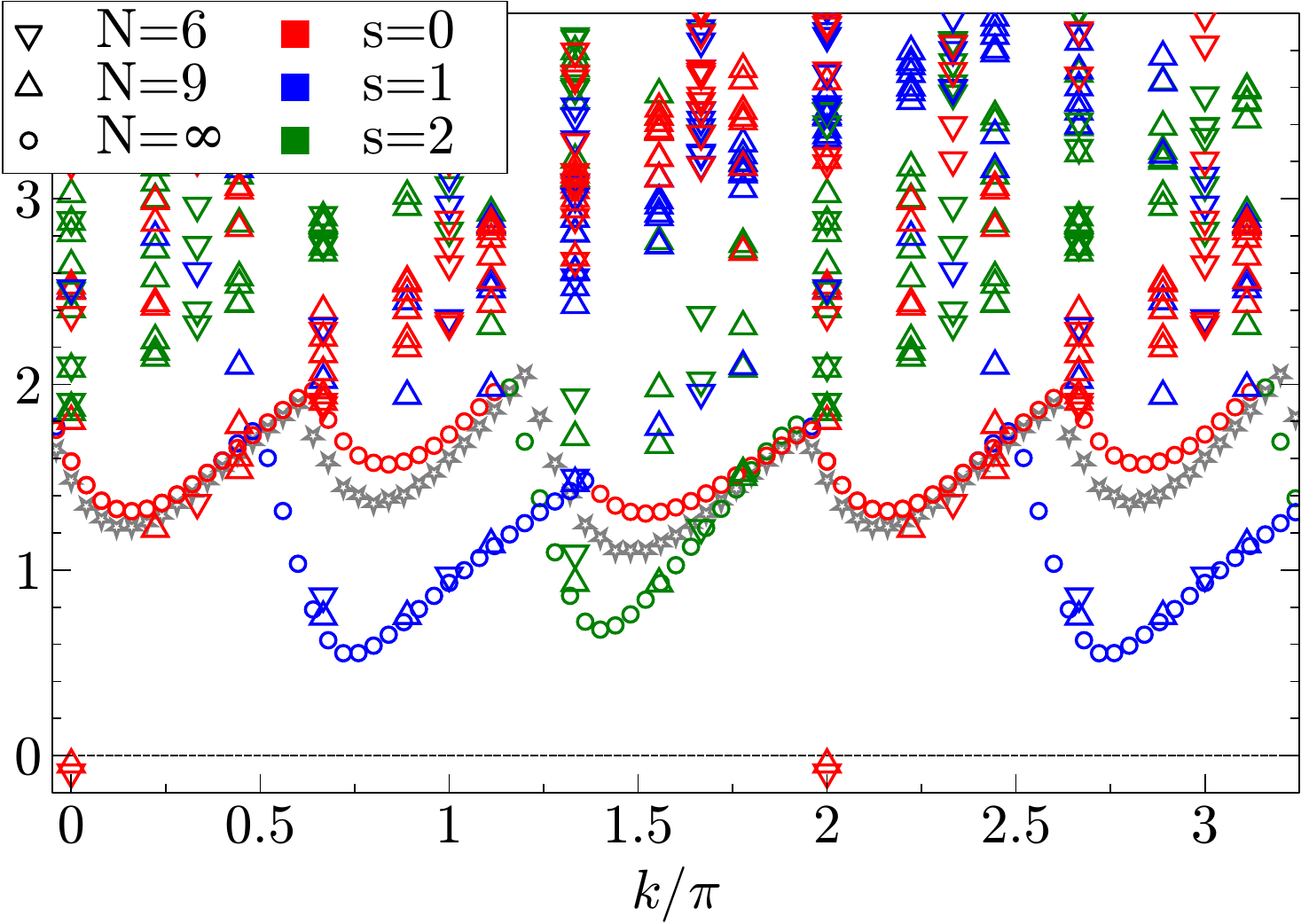}
\caption{%
Entanglement spectrum of the $\mathrm{SU}(3)$ model, computed from the
iMPS fixed point exactly for $N=6,9$ and using an iMPS excitation ansatz for
$N=\infty$, labelled by $\mathbb Z_3$ symmetry sectors $s=0,1,2$. 
Finite $N$ spectra are trace-normalized per sector, and shifted by the
leading eigenvalue in the thermodynamic limit.
The gray stars indicate the lower edge of the two-particle continuum computed
from the $s=1$ and $s=2$ one-particle branches for $N=\infty$. The
interactions between the particles determine the finite-size energies of
these two-particle states; the fact that for $N=6,9$ we only see the
continuum around momentum zero indicates that the interactions are
attractive only in that region.  The fact that the $N=\infty$
quasiparticle-ansatz in that region yields results close to the
two-particle edge (as compared to the other two edges) confirms this
picture.}
\label{fig:entspec}
\end{figure}

\begin{figure}[t]
\includegraphics[width=\columnwidth]{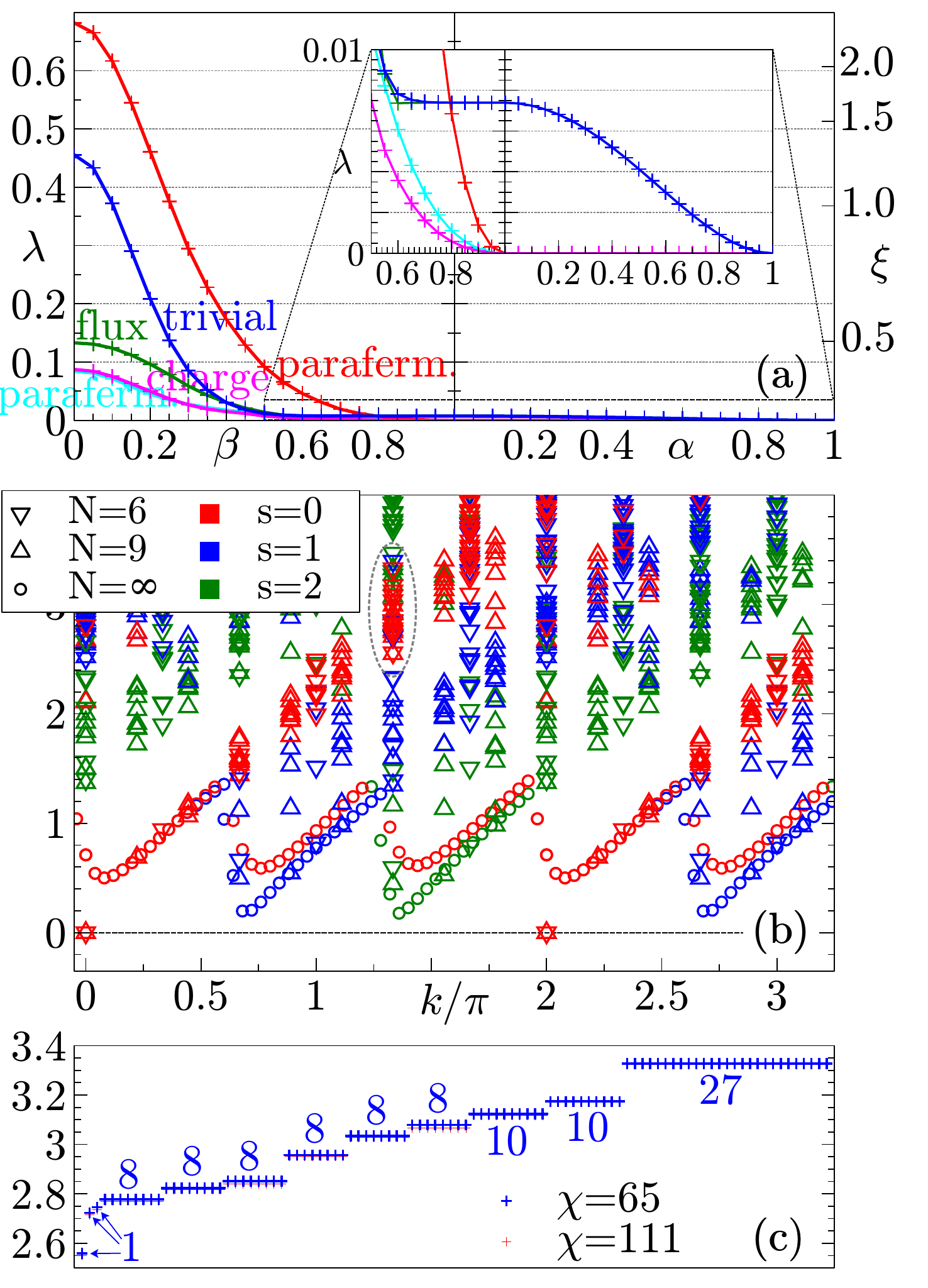}
\caption{The $\tb$-enhanced model. (a) Correlation functions
(cf.~Fig.~\ref{fig:corrlength}). (b) Entanglement spectrum
(cf.~Fig.~\ref{fig:entspec}).  Here, the finite $N$ data in the trivial
sector is scaled to leading eigenvalue $0$, and the relative normalization
of the other sectors is shifted by a factor $1.5$ such as to match the
$N=\infty$ data. 
Just as in Fig.~\ref{fig:entspec}, we can see that the finite-size data
only contains the momentum zero two-particle edge. This, together with the
counting, suggests that the states on this edge will evolve towards an
elementary chiral mode when the state approaches the critical point.
(c) The multiplet of lowest eigenvalues in the ES at $k=4\times(2\pi/N)$,
$N=6$ (ellipse in (b)), computed using two iMPS approximations with
$\chi=65,111$.  We perfectly recover the $\mathrm{SU}(3)_1$
counting~\cite{difrancesco:CFTbook} without having imposed any
symmetries. We also observe that the ES only depends very weakly on the
iMPS bond dimension $\chi$.}
\label{fig:W3b}
\end{figure}

The fact that our wavefunction transforms chirally under the lattice
symmetry raises the question whether it might display a chiral
ES (i.e., an ES described by a
chiral CFT).  We have studied the ES of the model at
the $\mathrm{SU}(3)$ point using an iMPS fixed point of the transfer
operator with $\chi=59$, using (\emph{i}) exact diagonalization on
cylinders up to circumference $N=9$ (which reveals a $3$-periodicity),
and (\emph{ii}) an iMPS excitation ansatz in the thermodynamic limit
($N=\infty)$~\cite{haegeman:shadows,haegeman:medley}.  The results for
$N=6,9,\infty$ are shown in Fig.~\ref{fig:entspec}, labeled by their
$\mathbb Z_3$ sector. On
the one hand, one can clearly see that the ES breaks
time-reversal and exhibits clear right-moving modes for all three
branches.  On the other hand, it also becomes clear from the $N=9$ and
$N=\infty$ data that the ES is gapped, with a much
steeper left-moving mode. (This is confirmed by computing the momentum
polarization~\cite{tu:momentum-polarization} from the
iMPS~\cite{poilblanc:kl-peps-2}.) These features become more pronounced when
considering the ES for the $\tb$-enhanced model (Fig.~\ref{fig:W3b}b).
In that case, the lowest branches become clearly
separated from the rest of the spectrum, allowing to perform a counting of
the spin multiplets.  The results in the trivial sector are in full
agreement with the counting and spin multiplet structure for a chiral
$\mathrm{SU}(3)_1$ CFT for at least the first $5$ levels
(Fig.~\ref{fig:W3b}c).  Nevertheless, inclusion of the $N=9$ and
$N=\infty$ data strongly indicates that the ES is still gapped.

\section{Conclusions}

In conclusion, we have presented a PEPS model which realizes a
$\mathrm{SU}(3)$ TSL with $\mathbb Z_3$ topological
order, and comprehensively characterized its
topological and entanglement properties. A number of questions remain:
First, it would be interesting to see whether twisted versions of
the model can be built without breaking lattice symmetries; this seems
plausible given the unbiased nature of the mapping to the 
topological $\mathbb Z_3$ model. Second, the model transforms chirally
under reflection, which seems unavoidable since different singlets in
$\ket\tau$ transform differently under reflection (and removing either kind
results in a critical or trivial model);  it would thus be interesting to
clarify whether there are fundamental obstacles for constructing a
trivially transforming $\mathrm{SU}(3)$ spin liquid.
Third, the connection to realistic systems is up for further study. 
In the case of $\mathrm{SU}(2)$, the motivation for considering RVBs
originates in the search for spin liquids in (frustrated) Heisenberg
models, and the question arises how our PEPS construction can serve a
similar purpose for $\mathrm{SU}(N)$. Although the simplest
$\mathrm{SU}(N)$ Heisenberg models seem to form symmetry broken simplex
solids~\cite{corboz:suN-heisenberg-simplex-solids}, one can look for
additional spin interactions that stabilize spin liquids of the form that
we presented~\footnote{Note that similarly, also AKLT-type constructions
for higher $\mathrm{SU}(N)$ exhibit symmetry
breaking~\cite{arovas:suN-simplex-solids}, so that it is an interesting
question whether our construction for higher $N$ (at least $N=4$ is
straightforward) will eventually cease to give a spin liquid.}, or try to
invoke the connection between PEPS and two-body Hamiltonians through
perturbative constructions~\cite{brell:tcode-perturbation-gadget}.  An
alternative route towards experimental implementations could be to devise
schemes for the controlled dynamical preparation of such wavefunctions in
e.g.\ optical lattices, using for instance the connection between PEPS and
perturbative constructions.  
Finally, a natural follow-up question is to investigate whether the model
can be further deformed to acquire a chiral $\mathrm{SU}(3)_1$ ES, e.g.\
by further increasing the weight of the $\tb$, and whether this happens at
the transition into the trivial phase, as well as to to better understand
its relation to other chiral models with $\mathrm{SU}(3)_1$
ES~\cite{wu:su3-chiral-kagome-SL,nataf:suN-chiral-SL}.

\subsection*{Acknowledgements}

We acknowledge helpful discussions with I.~Cirac, A.~Hackenbroich,
D.~Poilblanc,  A.~Sterdyniak, and H.-H.~Tu.  This work has been supported
by the EU through the ERC Starting Grant WASCOSYS (No.~636201), and by the
Flemish Research Foundation.

\section*{Appendix: The $\tb$-enhanced model}

In this Appendix, we describe the construction of the
$\tb$-enhanced model and provide numerical results on its correlation
functions, entanglement spectrum, and CFT counting of the ES.

The model is constructed in complete analogy to the original model
introduced in the main text, with the difference that the interpolation
$\mathcal P_{(\alpha,1)}$ is constructed such as to keep the total weight
of the $\tb\otimes \tb$ subspace constant along the interpolation (in
which we continuously remove the $\bm{\bar{6}}$ component in
$\tb\otimes\tb=3\oplus\bm{\bar{6}}$). Specifically, this amounts to
replacing Eq.~(\ref{eq:first-P-interpolation}) by 
\begin{equation}
    \tag{$\ref{eq:first-P-interpolation}^\prime$}
\mathcal P'_{(\alpha,1)}= \Big[\openone_{\mathcal H_\omega}
    \oplus \openone_{\mathcal H_{\bar\omega}}
    \oplus \big(\alpha \openone_{\mathcal H_1} + 
    (\sqrt{3-2\alpha^2}-\alpha) \Pi_{\mathcal H_1^{\bm 3}}\big)\Big]\, P_\perp\ .
\end{equation}
Correspondingly, $\mathcal P_{(0,\beta)}$,
Eq.~\eqref{eq:second-P-interpolation}, needs to be modified by replacing
$\mathcal P_{(0,1)}$ on the r.h.s.\ by 
\[
\mathcal P'_{(0,1)}= \Big[\openone_{\mathcal H_\omega}
    \oplus \openone_{\mathcal H_{\bar\omega}}
    \oplus \sqrt{3}\, \Pi_{\mathcal H_1^{\bm 3}}\Big]\, P_\perp\ .
\]

The analysis of the correlations, shown in Fig.~\ref{fig:W3b}a, reveals
that the modification indeed succeeds in decreasing the correlation length
significantly during the $\alpha$-interpolation.  On the other hand,
we find that as we approach the $\mathrm{SU}(3)$ invariant point along the
$\beta$-interpolation, the effect is reversed, and the final correlation
length at the $\mathrm{SU}(3)$ invariant point $\beta=0$ is in fact larger
than for the original model, $\xi\approx2.64$. This can be understood
from the fact that the deformation increases the weight of the
$\tb\otimes\tb\otimes\tb$ configuration in $\tau$, which will ultimately
drive the system through a transition into a trivial phase.

The entanglement spectrum of the $\tb$-enhanced model, shown in
Fig.~\ref{fig:W3b}b, exhibits a much more pronounced difference between the
left- and right-propagating modes, suggesting that the model might
eventually become chiral when further increasing the weight of the $\tb$
configurations, possibly at the phase transition. The clearly separated
right-moving branches allows for the identification and counting of the
spin multiplets for a given momentum $k=2\pi n/N$, which we find to
perfectly match the counting of the $\mathrm{SU}(3)_1$ CFT.  For
illustration, the $n=4$ multiplet is shown in Fig.~\ref{fig:W3b}c.


\begin{thebibliography}{36}%
\makeatletter
\providecommand \@ifxundefined [1]{%
 \@ifx{#1\undefined}
}%
\providecommand \@ifnum [1]{%
 \ifnum #1\expandafter \@firstoftwo
 \else \expandafter \@secondoftwo
 \fi
}%
\providecommand \@ifx [1]{%
 \ifx #1\expandafter \@firstoftwo
 \else \expandafter \@secondoftwo
 \fi
}%
\providecommand \natexlab [1]{#1}%
\providecommand \enquote  [1]{``#1''}%
\providecommand \bibnamefont  [1]{#1}%
\providecommand \bibfnamefont [1]{#1}%
\providecommand \citenamefont [1]{#1}%
\providecommand \href@noop [0]{\@secondoftwo}%
\providecommand \href [0]{\begingroup \@sanitize@url \@href}%
\providecommand \@href[1]{\@@startlink{#1}\@@href}%
\providecommand \@@href[1]{\endgroup#1\@@endlink}%
\providecommand \@sanitize@url [0]{\catcode `\\12\catcode `\$12\catcode
  `\&12\catcode `\#12\catcode `\^12\catcode `\_12\catcode `\%12\relax}%
\providecommand \@@startlink[1]{}%
\providecommand \@@endlink[0]{}%
\providecommand \url  [0]{\begingroup\@sanitize@url \@url }%
\providecommand \@url [1]{\endgroup\@href {#1}{\urlprefix }}%
\providecommand \urlprefix  [0]{URL }%
\providecommand \Eprint [0]{\href }%
\providecommand \doibase [0]{http://dx.doi.org/}%
\providecommand \selectlanguage [0]{\@gobble}%
\providecommand \bibinfo  [0]{\@secondoftwo}%
\providecommand \bibfield  [0]{\@secondoftwo}%
\providecommand \translation [1]{[#1]}%
\providecommand \BibitemOpen [0]{}%
\providecommand \bibitemStop [0]{}%
\providecommand \bibitemNoStop [0]{.\EOS\space}%
\providecommand \EOS [0]{\spacefactor3000\relax}%
\providecommand \BibitemShut  [1]{\csname bibitem#1\endcsname}%
\let\auto@bib@innerbib\@empty
%</preamble>
\bibitem [{\citenamefont {Anderson}(1987)}]{anderson:rvb-highTC}%
  \BibitemOpen
  \bibfield  {author} {\bibinfo {author} {\bibfnamefont {P.~W.}\ \bibnamefont
  {Anderson}},\ }\href@noop {} {\bibfield  {journal} {\bibinfo  {journal}
  {Science}\ }\textbf {\bibinfo {volume} {235}},\ \bibinfo {pages} {1196}
  (\bibinfo {year} {1987})}\BibitemShut {NoStop}%
\bibitem [{\citenamefont {Moessner}\ and\ \citenamefont
  {Sondhi}(2001)}]{moessner:dimer-triangular}%
  \BibitemOpen
  \bibfield  {author} {\bibinfo {author} {\bibfnamefont {R.}~\bibnamefont
  {Moessner}}\ and\ \bibinfo {author} {\bibfnamefont {S.~L.}\ \bibnamefont
  {Sondhi}},\ }\href@noop {} {\bibfield  {journal} {\bibinfo  {journal} {Phys.
  Rev. Lett.}\ }\textbf {\bibinfo {volume} {86}},\ \bibinfo {pages} {1881}
  (\bibinfo {year} {2001})},\ \Eprint {http://arxiv.org/abs/cond-mat/0007378}
  {cond-mat/0007378} \BibitemShut {NoStop}%
\bibitem [{\citenamefont {Misguich}\ \emph {et~al.}(2002)\citenamefont
  {Misguich}, \citenamefont {Serban},\ and\ \citenamefont
  {Pasquier}}]{misguich:dimer-kagome}%
  \BibitemOpen
  \bibfield  {author} {\bibinfo {author} {\bibfnamefont {G.}~\bibnamefont
  {Misguich}}, \bibinfo {author} {\bibfnamefont {D.}~\bibnamefont {Serban}}, \
  and\ \bibinfo {author} {\bibfnamefont {V.}~\bibnamefont {Pasquier}},\
  }\href@noop {} {\bibfield  {journal} {\bibinfo  {journal} {Phys. Rev. Lett.}\
  }\textbf {\bibinfo {volume} {89}},\ \bibinfo {pages} {137202} (\bibinfo
  {year} {2002})},\ \Eprint {http://arxiv.org/abs/cond-mat/0204428}
  {cond-mat/0204428} \BibitemShut {NoStop}%
\bibitem [{\citenamefont {Verstraete}\ and\ \citenamefont
  {Cirac}(2004)}]{verstraete:mbc-peps}%
  \BibitemOpen
  \bibfield  {author} {\bibinfo {author} {\bibfnamefont {F.}~\bibnamefont
  {Verstraete}}\ and\ \bibinfo {author} {\bibfnamefont {J.~I.}\ \bibnamefont
  {Cirac}},\ }\href@noop {} {\bibfield  {journal} {\bibinfo  {journal}
  {Phys.~Rev.~A}\ }\textbf {\bibinfo {volume} {70}},\ \bibinfo {pages} {060302}
  (\bibinfo {year} {2004})},\ \Eprint {http://arxiv.org/abs/quant-ph/0311130}
  {quant-ph/0311130} \BibitemShut {NoStop}%
\bibitem [{\citenamefont {Schuch}\ \emph {et~al.}(2012)\citenamefont {Schuch},
  \citenamefont {Poilblanc}, \citenamefont {Cirac},\ and\ \citenamefont
  {P{\'e}rez-Garc{\'\i}a}}]{schuch:rvb-kagome}%
  \BibitemOpen
  \bibfield  {author} {\bibinfo {author} {\bibfnamefont {N.}~\bibnamefont
  {Schuch}}, \bibinfo {author} {\bibfnamefont {D.}~\bibnamefont {Poilblanc}},
  \bibinfo {author} {\bibfnamefont {J.~I.}\ \bibnamefont {Cirac}}, \ and\
  \bibinfo {author} {\bibfnamefont {D.}~\bibnamefont {P{\'e}rez-Garc{\'\i}a}},\
  }\href@noop {} {\bibfield  {journal} {\bibinfo  {journal} {Phys. Rev. B}\
  }\textbf {\bibinfo {volume} {86}},\ \bibinfo {pages} {115108} (\bibinfo
  {year} {2012})},\ \Eprint {http://arxiv.org/abs/arXiv:1203.4816}
  {arXiv:1203.4816} \BibitemShut {NoStop}%
\bibitem [{\citenamefont {Hackenbroich}\ \emph {et~al.}(2018)\citenamefont
  {Hackenbroich}, \citenamefont {Sterdyniak},\ and\ \citenamefont
  {Schuch}}]{hackenbroich:chiral-su2}%
  \BibitemOpen
  \bibfield  {author} {\bibinfo {author} {\bibfnamefont {A.}~\bibnamefont
  {Hackenbroich}}, \bibinfo {author} {\bibfnamefont {A.}~\bibnamefont
  {Sterdyniak}}, \ and\ \bibinfo {author} {\bibfnamefont {N.}~\bibnamefont
  {Schuch}},\ }\href@noop {} {\  (\bibinfo {year} {2018})},\ \Eprint
  {http://arxiv.org/abs/arXiv:1805.04531} {arXiv:1805.04531} \BibitemShut
  {NoStop}%
\bibitem [{\citenamefont {{Schuch}}\ \emph {et~al.}(2010)\citenamefont
  {{Schuch}}, \citenamefont {{Cirac}},\ and\ \citenamefont
  {{P{\'e}rez-Garc{\'{\i}}a}}}]{schuch:peps-sym}%
  \BibitemOpen
  \bibfield  {author} {\bibinfo {author} {\bibfnamefont {N.}~\bibnamefont
  {{Schuch}}}, \bibinfo {author} {\bibfnamefont {I.}~\bibnamefont {{Cirac}}}, \
  and\ \bibinfo {author} {\bibfnamefont {D.}~\bibnamefont
  {{P{\'e}rez-Garc{\'{\i}}a}}},\ }\href {\doibase 10.1016/j.aop.2010.05.008}
  {\bibfield  {journal} {\bibinfo  {journal} {Ann. Phys.}\ }\textbf {\bibinfo
  {volume} {325}},\ \bibinfo {pages} {2153} (\bibinfo {year} {2010})},\ \Eprint
  {http://arxiv.org/abs/arXiv:1001.3807} {arXiv:1001.3807} \BibitemShut
  {NoStop}%
\bibitem [{\citenamefont {Iqbal}\ \emph {et~al.}(2014)\citenamefont {Iqbal},
  \citenamefont {Poilblanc},\ and\ \citenamefont
  {Schuch}}]{iqbal:semionic-rvb}%
  \BibitemOpen
  \bibfield  {author} {\bibinfo {author} {\bibfnamefont {M.}~\bibnamefont
  {Iqbal}}, \bibinfo {author} {\bibfnamefont {D.}~\bibnamefont {Poilblanc}}, \
  and\ \bibinfo {author} {\bibfnamefont {N.}~\bibnamefont {Schuch}},\
  }\href@noop {} {\bibfield  {journal} {\bibinfo  {journal} {Phys. Rev. B}\
  }\textbf {\bibinfo {volume} {90}},\ \bibinfo {pages} {115129} (\bibinfo
  {year} {2014})},\ \Eprint {http://arxiv.org/abs/1407.7773} {1407.7773}
  \BibitemShut {NoStop}%
\bibitem [{\citenamefont {Qi}\ \emph {et~al.}(2015)\citenamefont {Qi},
  \citenamefont {Gu},\ and\ \citenamefont {Yao}}]{qi:semionic-dimer-models}%
  \BibitemOpen
  \bibfield  {author} {\bibinfo {author} {\bibfnamefont {Y.}~\bibnamefont
  {Qi}}, \bibinfo {author} {\bibfnamefont {Z.-C.}\ \bibnamefont {Gu}}, \ and\
  \bibinfo {author} {\bibfnamefont {H.}~\bibnamefont {Yao}},\ }\href {\doibase
  10.1103/PhysRevB.92.155105} {\bibfield  {journal} {\bibinfo  {journal} {Phys.
  Rev. B}\ }\textbf {\bibinfo {volume} {92}},\ \bibinfo {pages} {155105}
  (\bibinfo {year} {2015})},\ \Eprint {http://arxiv.org/abs/arXiv:1406.6364}
  {arXiv:1406.6364} \BibitemShut {NoStop}%
\bibitem [{\citenamefont {Buerschaper}\ \emph {et~al.}(2014)\citenamefont
  {Buerschaper}, \citenamefont {Morampudi},\ and\ \citenamefont
  {Pollmann}}]{buerschaper:semionic-dimermodel}%
  \BibitemOpen
  \bibfield  {author} {\bibinfo {author} {\bibfnamefont {O.}~\bibnamefont
  {Buerschaper}}, \bibinfo {author} {\bibfnamefont {S.~C.}\ \bibnamefont
  {Morampudi}}, \ and\ \bibinfo {author} {\bibfnamefont {F.}~\bibnamefont
  {Pollmann}},\ }\href {\doibase 10.1103/PhysRevB.90.195148} {\bibfield
  {journal} {\bibinfo  {journal} {Phys. Rev. B}\ }\textbf {\bibinfo {volume}
  {90}},\ \bibinfo {pages} {195148} (\bibinfo {year} {2014})},\ \Eprint
  {http://arxiv.org/abs/arXiv:1407.8521} {arXiv:1407.8521} \BibitemShut
  {NoStop}%
\bibitem [{\citenamefont {Zaletel}\ and\ \citenamefont
  {Vishwanath}(2015)}]{zaletel:su2-semion}%
  \BibitemOpen
  \bibfield  {author} {\bibinfo {author} {\bibfnamefont {M.~P.}\ \bibnamefont
  {Zaletel}}\ and\ \bibinfo {author} {\bibfnamefont {A.}~\bibnamefont
  {Vishwanath}},\ }\href {\doibase 10.1103/PhysRevLett.114.077201} {\bibfield
  {journal} {\bibinfo  {journal} {Phys.\ Rev.\ Lett.}\ }\textbf {\bibinfo
  {volume} {114}},\ \bibinfo {pages} {077201} (\bibinfo {year} {2015})},\
  \Eprint {http://arxiv.org/abs/arXiv:1410.2894} {arXiv:1410.2894} \BibitemShut
  {NoStop}%
\bibitem [{\citenamefont {Gorshkov}\ \emph {et~al.}(2010)\citenamefont
  {Gorshkov}, \citenamefont {Hermele}, \citenamefont {Gurarie}, \citenamefont
  {Xu}, \citenamefont {Julienne}, \citenamefont {Ye}, \citenamefont {Zoller},
  \citenamefont {Demler}, \citenamefont {Lukin},\ and\ \citenamefont
  {Rey}}]{gorshkov:suN-optical-lattice}%
  \BibitemOpen
  \bibfield  {author} {\bibinfo {author} {\bibfnamefont {A.~V.}\ \bibnamefont
  {Gorshkov}}, \bibinfo {author} {\bibfnamefont {M.}~\bibnamefont {Hermele}},
  \bibinfo {author} {\bibfnamefont {V.}~\bibnamefont {Gurarie}}, \bibinfo
  {author} {\bibfnamefont {C.}~\bibnamefont {Xu}}, \bibinfo {author}
  {\bibfnamefont {P.~S.}\ \bibnamefont {Julienne}}, \bibinfo {author}
  {\bibfnamefont {J.}~\bibnamefont {Ye}}, \bibinfo {author} {\bibfnamefont
  {P.}~\bibnamefont {Zoller}}, \bibinfo {author} {\bibfnamefont
  {E.}~\bibnamefont {Demler}}, \bibinfo {author} {\bibfnamefont {M.~D.}\
  \bibnamefont {Lukin}}, \ and\ \bibinfo {author} {\bibfnamefont {A.~M.}\
  \bibnamefont {Rey}},\ }\href {\doibase 10.1038/NPHYS1535} {\bibfield
  {journal} {\bibinfo  {journal} {Nature Phys.}\ }\textbf {\bibinfo {volume}
  {6}},\ \bibinfo {pages} {289} (\bibinfo {year} {2010})},\ \Eprint
  {http://arxiv.org/abs/arXiv:0905.2610} {arXiv:0905.2610} \BibitemShut
  {NoStop}%
\bibitem [{\citenamefont {Scazza}\ \emph {et~al.}(2014)\citenamefont {Scazza},
  \citenamefont {Hofrichter}, \citenamefont {Höfer}, \citenamefont {Groot},
  \citenamefont {Bloch},\ and\ \citenamefont
  {Fölling}}]{scazza:suN-optical-lattice}%
  \BibitemOpen
  \bibfield  {author} {\bibinfo {author} {\bibfnamefont {F.}~\bibnamefont
  {Scazza}}, \bibinfo {author} {\bibfnamefont {C.}~\bibnamefont {Hofrichter}},
  \bibinfo {author} {\bibfnamefont {M.}~\bibnamefont {Höfer}}, \bibinfo
  {author} {\bibfnamefont {P.~C.~D.}\ \bibnamefont {Groot}}, \bibinfo {author}
  {\bibfnamefont {I.}~\bibnamefont {Bloch}}, \ and\ \bibinfo {author}
  {\bibfnamefont {S.}~\bibnamefont {Fölling}},\ }\href {\doibase
  10.1038/nphys3061} {\bibfield  {journal} {\bibinfo  {journal} {Nature Phys.}\
  }\textbf {\bibinfo {volume} {10}},\ \bibinfo {pages} {779} (\bibinfo {year}
  {2014})},\ \Eprint {http://arxiv.org/abs/arXiv:1403.4761} {arXiv:1403.4761}
  \BibitemShut {NoStop}%
\bibitem [{\citenamefont {Zhang}\ \emph {et~al.}(2014)\citenamefont {Zhang},
  \citenamefont {Bishof}, \citenamefont {Bromley}, \citenamefont {Kraus},
  \citenamefont {Safronova}, \citenamefont {Zoller}, \citenamefont {Rey},\ and\
  \citenamefont {Ye}}]{zhang:suN-experiment}%
  \BibitemOpen
  \bibfield  {author} {\bibinfo {author} {\bibfnamefont {X.}~\bibnamefont
  {Zhang}}, \bibinfo {author} {\bibfnamefont {M.}~\bibnamefont {Bishof}},
  \bibinfo {author} {\bibfnamefont {S.~L.}\ \bibnamefont {Bromley}}, \bibinfo
  {author} {\bibfnamefont {C.~V.}\ \bibnamefont {Kraus}}, \bibinfo {author}
  {\bibfnamefont {M.~S.}\ \bibnamefont {Safronova}}, \bibinfo {author}
  {\bibfnamefont {P.}~\bibnamefont {Zoller}}, \bibinfo {author} {\bibfnamefont
  {A.~M.}\ \bibnamefont {Rey}}, \ and\ \bibinfo {author} {\bibfnamefont
  {J.}~\bibnamefont {Ye}},\ }\href {\doibase 10.1126/science.1254978}
  {\bibfield  {journal} {\bibinfo  {journal} {Science}\ }\textbf {\bibinfo
  {volume} {345}},\ \bibinfo {pages} {1467} (\bibinfo {year} {2014})},\ \Eprint
  {http://arxiv.org/abs/arXiv:1403.2964} {arXiv:1403.2964} \BibitemShut
  {NoStop}%
\bibitem [{\citenamefont {Hermele}\ and\ \citenamefont
  {Gurarie}(2011)}]{hermele:suN-liquids}%
  \BibitemOpen
  \bibfield  {author} {\bibinfo {author} {\bibfnamefont {M.}~\bibnamefont
  {Hermele}}\ and\ \bibinfo {author} {\bibfnamefont {V.}~\bibnamefont
  {Gurarie}},\ }\href {\doibase 10.1103/PhysRevB.84.174441} {\bibfield
  {journal} {\bibinfo  {journal} {Phys.\ Rev.\ B}\ }\textbf {\bibinfo {volume}
  {84}},\ \bibinfo {pages} {174441} (\bibinfo {year} {2011})},\ \Eprint
  {http://arxiv.org/abs/arXiv:1108.3862} {arXiv:1108.3862} \BibitemShut
  {NoStop}%
\bibitem [{\citenamefont {Nataf}\ \emph {et~al.}(2016)\citenamefont {Nataf},
  \citenamefont {Lajko}, \citenamefont {Wietek}, \citenamefont {Penc},
  \citenamefont {Mila},\ and\ \citenamefont {Laeuchli}}]{nataf:suN-chiral-SL}%
  \BibitemOpen
  \bibfield  {author} {\bibinfo {author} {\bibfnamefont {P.}~\bibnamefont
  {Nataf}}, \bibinfo {author} {\bibfnamefont {M.}~\bibnamefont {Lajko}},
  \bibinfo {author} {\bibfnamefont {A.}~\bibnamefont {Wietek}}, \bibinfo
  {author} {\bibfnamefont {K.}~\bibnamefont {Penc}}, \bibinfo {author}
  {\bibfnamefont {F.}~\bibnamefont {Mila}}, \ and\ \bibinfo {author}
  {\bibfnamefont {A.~M.}\ \bibnamefont {Laeuchli}},\ }\href {\doibase
  10.1103/PhysRevLett.117.167202} {\bibfield  {journal} {\bibinfo  {journal}
  {Phys.\ Rev.\ Lett.}\ }\textbf {\bibinfo {volume} {117}},\ \bibinfo {pages}
  {167202} (\bibinfo {year} {2016})},\ \Eprint
  {http://arxiv.org/abs/arXiv:1601.00958} {arXiv:1601.00958} \BibitemShut
  {NoStop}%
\bibitem [{Note1()}]{Note1}%
  \BibitemOpen
  \bibinfo {note} {Observe that our model is not a ``resonating trimer state''
  as e.g.\ in Ref.~\cite {lee:resonating-trimer-state}, since projecting
  $\protect \mathaccentV {bar}016{\protect \bm {3}}\otimes \protect
  \mathaccentV {bar}016{\protect \bm {3}}$ creates large entangled clusters
  (e.g.\ for the vaccuum of the $\protect \mathbb Z_3$ model). Placing the
  model of Ref.~\cite {lee:resonating-trimer-state} on the kagome lattice in
  fact yields a variant of our model with a modified \protect \text
  {$\delimiter "026A30C \tau \delimiter "526930B $} which we found to be in a
  trivial phase.}\BibitemShut {Stop}%
\bibitem [{Note2()}]{Note2}%
  \BibitemOpen
  \bibinfo {note} {Using $\protect \mathrm {U}(1)$ conservation and rewriting
  $M^\dagger M$ as a product of $6$ matrices each depending on the virtual
  ket+bra indices on one tip of the star; note that it is sufficient to check
  this for $\protect \mathcal P_{(0,1)}$, since for $\beta >0$, $\protect
  \mathcal P_{(\alpha ,\beta )}$ can be mapped to $\protect \mathcal P_{(0,1)}$
  by acting on the physical system.}\BibitemShut {Stop}%
\bibitem [{Note3()}]{Note3}%
  \BibitemOpen
  \bibinfo {note} {The latter point might still be $\protect \mathbb
  Z_3$-injective, but this would require checking larger patches which seems
  numerically prohibitive.}\BibitemShut {Stop}%
\bibitem [{Note4()}]{Note4}%
  \BibitemOpen
  \bibinfo {note} {The existence of the limit is guaranteed by the fact that
  the parent Hamiltonian is constructed as the projector onto the
  orthocomplement of the image of the PEPS on two overlapping stars, which is a
  polynomial (and thus analytic) map in $\beta $, together with the fact that
  eigenspace projectors of analytic hermitian matrices are analytic themselves
  (\cite [Thm.~6.1]{kato:pert-of-linear-operators}).}\BibitemShut {Stop}%
\bibitem [{\citenamefont {{Haegeman}}\ and\ \citenamefont
  {{Verstraete}}(2017)}]{haegeman:medley}%
  \BibitemOpen
  \bibfield  {author} {\bibinfo {author} {\bibfnamefont {J.}~\bibnamefont
  {{Haegeman}}}\ and\ \bibinfo {author} {\bibfnamefont {F.}~\bibnamefont
  {{Verstraete}}},\ }\href {\doibase 10.1146/annurev-conmatphys-031016-025507}
  {\bibfield  {journal} {\bibinfo  {journal} {Annual Review of Condensed Matter
  Physics}\ }\textbf {\bibinfo {volume} {8}},\ \bibinfo {pages} {355} (\bibinfo
  {year} {2017})},\ \Eprint {http://arxiv.org/abs/arXiv:1611.08519}
  {arXiv:1611.08519} \BibitemShut {NoStop}%
\bibitem [{\citenamefont {Iqbal}\ \emph {et~al.}(2018)\citenamefont {Iqbal},
  \citenamefont {Duivenvoorden},\ and\ \citenamefont
  {Schuch}}]{iqbal:z4-phasetrans}%
  \BibitemOpen
  \bibfield  {author} {\bibinfo {author} {\bibfnamefont {M.}~\bibnamefont
  {Iqbal}}, \bibinfo {author} {\bibfnamefont {K.}~\bibnamefont
  {Duivenvoorden}}, \ and\ \bibinfo {author} {\bibfnamefont {N.}~\bibnamefont
  {Schuch}},\ }\href@noop {} {\bibfield  {journal} {\bibinfo  {journal} {Phys.
  Rev. B}\ }\textbf {\bibinfo {volume} {97}},\ \bibinfo {pages} {195124}
  (\bibinfo {year} {2018})},\ \Eprint {http://arxiv.org/abs/arXiv:1712.04021}
  {arXiv:1712.04021} \BibitemShut {NoStop}%
\bibitem [{Note5()}]{Note5}%
  \BibitemOpen
  \bibinfo {note} {The longest-ranging two-point correlations (corresponding to
  the trivial sector) with $\xi \approx 0.86$ are obtained for spin-spin
  correlations.}\BibitemShut {Stop}%
\bibitem [{\citenamefont {{Duivenvoorden}}\ \emph {et~al.}(2017)\citenamefont
  {{Duivenvoorden}}, \citenamefont {{Iqbal}}, \citenamefont {{Haegeman}},
  \citenamefont {{Verstraete}},\ and\ \citenamefont
  {{Schuch}}}]{duivenvoorden:anyon-condensation}%
  \BibitemOpen
  \bibfield  {author} {\bibinfo {author} {\bibfnamefont {K.}~\bibnamefont
  {{Duivenvoorden}}}, \bibinfo {author} {\bibfnamefont {M.}~\bibnamefont
  {{Iqbal}}}, \bibinfo {author} {\bibfnamefont {J.}~\bibnamefont {{Haegeman}}},
  \bibinfo {author} {\bibfnamefont {F.}~\bibnamefont {{Verstraete}}}, \ and\
  \bibinfo {author} {\bibfnamefont {N.}~\bibnamefont {{Schuch}}},\ }\href
  {\doibase 10.1103/PhysRevB.95.235119} {\bibfield  {journal} {\bibinfo
  {journal} {Phys. Rev. B}\ }\textbf {\bibinfo {volume} {95}},\ \bibinfo {eid}
  {235119} (\bibinfo {year} {2017})},\ \Eprint
  {http://arxiv.org/abs/1702.08469} {arXiv:1702.08469} \BibitemShut {NoStop}%
\bibitem [{Note6()}]{Note6}%
  \BibitemOpen
  \bibinfo {note} {Exact diagonalization of the transfer operator on small
  cylinders, the convergence of the iMPS method, and the smooth change of the
  leading eigenvalue and eigenvector of the transfer operator along the
  interpolation.}\BibitemShut {Stop}%
\bibitem [{\citenamefont {Haegeman}\ \emph {et~al.}(2015)\citenamefont
  {Haegeman}, \citenamefont {Zauner}, \citenamefont {Schuch},\ and\
  \citenamefont {Verstraete}}]{haegeman:shadows}%
  \BibitemOpen
  \bibfield  {author} {\bibinfo {author} {\bibfnamefont {J.}~\bibnamefont
  {Haegeman}}, \bibinfo {author} {\bibfnamefont {V.}~\bibnamefont {Zauner}},
  \bibinfo {author} {\bibfnamefont {N.}~\bibnamefont {Schuch}}, \ and\ \bibinfo
  {author} {\bibfnamefont {F.}~\bibnamefont {Verstraete}},\ }\href {\doibase
  doi:10.1038/ncomms9284} {\bibfield  {journal} {\bibinfo  {journal} {Nature
  Comm.}\ }\textbf {\bibinfo {volume} {6}},\ \bibinfo {pages} {8284} (\bibinfo
  {year} {2015})},\ \Eprint {http://arxiv.org/abs/arXiv:1410.5443}
  {arXiv:1410.5443} \BibitemShut {NoStop}%
\bibitem [{\citenamefont {Tu}\ \emph {et~al.}(2013)\citenamefont {Tu},
  \citenamefont {Zhang},\ and\ \citenamefont {Qi}}]{tu:momentum-polarization}%
  \BibitemOpen
  \bibfield  {author} {\bibinfo {author} {\bibfnamefont {H.-H.}\ \bibnamefont
  {Tu}}, \bibinfo {author} {\bibfnamefont {Y.}~\bibnamefont {Zhang}}, \ and\
  \bibinfo {author} {\bibfnamefont {X.-L.}\ \bibnamefont {Qi}},\ }\href
  {\doibase 10.1103/PhysRevB.88.195412} {\bibfield  {journal} {\bibinfo
  {journal} {Phys.\ Rev.\ B}\ }\textbf {\bibinfo {volume} {88}},\ \bibinfo
  {pages} {195412} (\bibinfo {year} {2013})},\ \Eprint
  {http://arxiv.org/abs/arXiv:1212.6951} {arXiv:1212.6951} \BibitemShut
  {NoStop}%
\bibitem [{\citenamefont {Poilblanc}\ \emph {et~al.}(2016)\citenamefont
  {Poilblanc}, \citenamefont {Schuch},\ and\ \citenamefont
  {Affleck}}]{poilblanc:kl-peps-2}%
  \BibitemOpen
  \bibfield  {author} {\bibinfo {author} {\bibfnamefont {D.}~\bibnamefont
  {Poilblanc}}, \bibinfo {author} {\bibfnamefont {N.}~\bibnamefont {Schuch}}, \
  and\ \bibinfo {author} {\bibfnamefont {I.}~\bibnamefont {Affleck}},\
  }\href@noop {} {\bibfield  {journal} {\bibinfo  {journal} {Phys. Rev. B}\
  }\textbf {\bibinfo {volume} {93}},\ \bibinfo {pages} {174414} (\bibinfo
  {year} {2016})},\ \Eprint {http://arxiv.org/abs/arXiv:1602.05969}
  {arXiv:1602.05969} \BibitemShut {NoStop}%
\bibitem [{\citenamefont {Corboz}\ \emph {et~al.}(2012)\citenamefont {Corboz},
  \citenamefont {Penc}, \citenamefont {Mila},\ and\ \citenamefont
  {Laeuchli}}]{corboz:suN-heisenberg-simplex-solids}%
  \BibitemOpen
  \bibfield  {author} {\bibinfo {author} {\bibfnamefont {P.}~\bibnamefont
  {Corboz}}, \bibinfo {author} {\bibfnamefont {K.}~\bibnamefont {Penc}},
  \bibinfo {author} {\bibfnamefont {F.}~\bibnamefont {Mila}}, \ and\ \bibinfo
  {author} {\bibfnamefont {A.~M.}\ \bibnamefont {Laeuchli}},\ }\href {\doibase
  10.1103/PhysRevB.86.041106} {\bibfield  {journal} {\bibinfo  {journal} {Phys.
  Rev. B}\ }\textbf {\bibinfo {volume} {86}},\ \bibinfo {pages} {041106}
  (\bibinfo {year} {2012})},\ \Eprint {http://arxiv.org/abs/arXiv:1204.6682}
  {arXiv:1204.6682} \BibitemShut {NoStop}%
\bibitem [{Note7()}]{Note7}%
  \BibitemOpen
  \bibinfo {note} {Note that similarly, also AKLT-type constructions for higher
  $\protect \mathrm {SU}(N)$ exhibit symmetry breaking~\cite
  {arovas:suN-simplex-solids}, so that it is an interesting question whether
  our construction for higher $N$ (at least $N=4$ is straightforward) will
  eventually cease to give a spin liquid.}\BibitemShut {Stop}%
\bibitem [{\citenamefont {Brell}\ \emph {et~al.}(2011)\citenamefont {Brell},
  \citenamefont {Flammia}, \citenamefont {Bartlett},\ and\ \citenamefont
  {Doherty}}]{brell:tcode-perturbation-gadget}%
  \BibitemOpen
  \bibfield  {author} {\bibinfo {author} {\bibfnamefont {C.~G.}\ \bibnamefont
  {Brell}}, \bibinfo {author} {\bibfnamefont {S.~T.}\ \bibnamefont {Flammia}},
  \bibinfo {author} {\bibfnamefont {S.~D.}\ \bibnamefont {Bartlett}}, \ and\
  \bibinfo {author} {\bibfnamefont {A.~C.}\ \bibnamefont {Doherty}},\ }\href
  {\doibase 10.1088/1367-2630/13/5/053039} {\bibfield  {journal} {\bibinfo
  {journal} {New J. Phys.}\ }\textbf {\bibinfo {volume} {13}},\ \bibinfo
  {pages} {053039} (\bibinfo {year} {2011})},\ \Eprint
  {http://arxiv.org/abs/arXiv:1011.1942} {arXiv:1011.1942} \BibitemShut
  {NoStop}%
\bibitem [{\citenamefont {Wu}\ and\ \citenamefont
  {Tu}(2016)}]{wu:su3-chiral-kagome-SL}%
  \BibitemOpen
  \bibfield  {author} {\bibinfo {author} {\bibfnamefont {Y.-H.}\ \bibnamefont
  {Wu}}\ and\ \bibinfo {author} {\bibfnamefont {H.-H.}\ \bibnamefont {Tu}},\
  }\href {\doibase 10.1103/PhysRevB.94.201113} {\bibfield  {journal} {\bibinfo
  {journal} {Phys. Rev. B}\ }\textbf {\bibinfo {volume} {94}},\ \bibinfo
  {pages} {201113} (\bibinfo {year} {2016})},\ \Eprint
  {http://arxiv.org/abs/arXiv:1601.02594} {arXiv:1601.02594} \BibitemShut
  {NoStop}%
\bibitem [{\citenamefont {Lee}\ \emph {et~al.}(2017)\citenamefont {Lee},
  \citenamefont {Oh}, \citenamefont {Han},\ and\ \citenamefont
  {Katsura}}]{lee:resonating-trimer-state}%
  \BibitemOpen
  \bibfield  {author} {\bibinfo {author} {\bibfnamefont {H.}~\bibnamefont
  {Lee}}, \bibinfo {author} {\bibfnamefont {Y.}~\bibnamefont {Oh}}, \bibinfo
  {author} {\bibfnamefont {J.~H.}\ \bibnamefont {Han}}, \ and\ \bibinfo
  {author} {\bibfnamefont {H.}~\bibnamefont {Katsura}},\ }\href {\doibase
  10.1103/PhysRevB.95.060413} {\bibfield  {journal} {\bibinfo  {journal} {Phys.
  Rev. B}\ }\textbf {\bibinfo {volume} {95}},\ \bibinfo {pages} {060413}
  (\bibinfo {year} {2017})},\ \Eprint {http://arxiv.org/abs/arXiv:1612.06899}
  {arXiv:1612.06899} \BibitemShut {NoStop}%
\bibitem [{\citenamefont {Kato}(1966)}]{kato:pert-of-linear-operators}%
  \BibitemOpen
  \bibfield  {author} {\bibinfo {author} {\bibfnamefont {T.}~\bibnamefont
  {Kato}},\ }\href {https://books.google.de/books?id=N\_HysgEACAAJ} {\emph
  {\bibinfo {title} {Perturbation theory for linear operators}}}\ (\bibinfo
  {publisher} {Springer Berlin Heidelberg},\ \bibinfo {year}
  {1966})\BibitemShut {NoStop}%
\bibitem [{\citenamefont {Arovas}(2008)}]{arovas:suN-simplex-solids}%
  \BibitemOpen
  \bibfield  {author} {\bibinfo {author} {\bibfnamefont {D.~P.}\ \bibnamefont
  {Arovas}},\ }\href {\doibase 10.1103/PhysRevB.77.104404} {\bibfield
  {journal} {\bibinfo  {journal} {Phys. Rev. B}\ }\textbf {\bibinfo {volume}
  {77}},\ \bibinfo {pages} {104404} (\bibinfo {year} {2008})},\ \Eprint
  {http://arxiv.org/abs/arXiv:0711.3921} {arXiv:0711.3921} \BibitemShut
  {NoStop}%
\bibitem [{\citenamefont {Di~Francesco}\ \emph {et~al.}(1997)\citenamefont
  {Di~Francesco}, \citenamefont {Mathieu},\ and\ \citenamefont
  {S{\'e}n{\'e}chal}}]{difrancesco:CFTbook}%
  \BibitemOpen
  \bibfield  {author} {\bibinfo {author} {\bibfnamefont {P.}~\bibnamefont
  {Di~Francesco}}, \bibinfo {author} {\bibfnamefont {P.}~\bibnamefont
  {Mathieu}}, \ and\ \bibinfo {author} {\bibfnamefont {D.}~\bibnamefont
  {S{\'e}n{\'e}chal}},\ }\href {https://books.google.de/books?id=keUrdME5rhIC}
  {\emph {\bibinfo {title} {Conformal Field Theory}}},\ Graduate Texts in
  Contemporary Physics\ (\bibinfo  {publisher} {Springer},\ \bibinfo {year}
  {1997})\BibitemShut {NoStop}%
\end{thebibliography}
\end{document}